\documentstyle[aps,epsf,array]{revtex}
\def\plb{{ \sl Phys. Lett. }}
\def\dky{D_{KY}}
\def\figno#1{Fig.~\ref{fig:#1}}
\def\eqnn#1{(\ref{#1})}
\def\lam#1{\lambda_{#1}}
\def\dmax{\Delta D_{max}}
\def\dmin{\Delta D_{min}}
\begin{document}
\title{
  Lyapunov Exponents, Transport and the
  Extensivity of Dimensional Loss
  }
\author{Kenichiro Aoki$^a$
  and Dimitri Kusnezov$^b$
  }
\address{$^a$Dept. of Physics, Keio University, {\it
    4---1---1} Hiyoshi, Kouhoku--ku, Yokohama 223--8521, Japan\\
  $^b$Center for Theoretical Physics, Sloane Physics Lab, Yale
  University, New Haven, CT\ 06520-8120} % \date{\today }
\maketitle

\begin{abstract}
  An explicit relation between the
  dimensional loss ($\Delta D$), entropy production and
  transport  is established under thermal gradients,
  relating the microscopic and macroscopic
  behaviors of the system.  The extensivity of  $\Delta D$
  in systems with bulk behavior follows from the
  relation. The maximum Lyapunov exponents in thermal equilibrium
  and $\Delta D$   in non-equilibrium depend on the
  choice of heat-baths, while their product is unique and macroscopic.
  Finite size corrections are also computed and all results are
  verified numerically.
\end{abstract}
% 05.45.Jn   High-dimensional chaos 
% 05.45.Pq   Numerical simulations of chaotic models
% 05.60.-k   Transport processes
% 05.60.Cd   Classical transport
% 05.70.Ln   Nonequilibrium and irreversible thermodynamics
% 05.45Df fractals
% 05.45Pq Num. sim. of chaos
% 44.10+i heat conduction
%\pacs{PACS numbers: 05.45.-a; 05.45.+b, 02.50.-r, 05.60.+w, 05.70.Ln
%  05.60.-k, 44.10.+i, 02.70.Ns}
%
Fractal structures in phase space have become the focus of attention
in the understanding of the relationship between microscopic dynamics
and macroscopic non-equilibrium physics\cite{attractor}.
In the escape-rate formalism, the properties of a fractal
repellor are known to govern the transport\cite{gas}.
In contrast, in boundary driven non-equilibrium steady states, the stationary
distribution is generally fractal, but the precise connection to transport
is not fully understood\cite{attractor}. This reduction in dimension,
$\Delta D$, has been argued to be related to transport\cite{cels,hp2,evans},
although the only precise understanding arises in the weak field limit of
the Lorentz gas\cite{cels}, which does not have clear generalization to
non-equilibrium many-body systems. The presence of fractals has also been
used to demonstrate how the second law of thermodynamics
is consistent with time-reversal invariant, deterministic dynamics\cite{hm}.
Many of these issues require the understanding of the Lyapunov
spectrum, whose analytic properties are known only in certain special
cases\cite{analy}.
In this letter we study general Hamiltonians coupled to two heat
baths at different temperatures at opposite sides of the system,
generating heat flow. We will see that the type of heat baths
chosen affect $\Delta D$ and the Lyapunov exponents, 
for the same boundary temperatures. We investigate
the meaning and the origins of extensivity of $\Delta D$
in systems with bulk behavior, including finite size corrections,
and relate them to transport. The new relations are verified
numerically. We systematically study the system not only
close to but also far from equilibrium, as well as the
dependence on the heat-baths themselves. While the extensivity of
dimensional loss has been disputed due to the incompatibility with
local equilibrium\cite{cels}, we will see that this is not an
issue.

In systems with bulk behavior, to say that $\Delta D$
behaves ``extensively'' means that $\Delta D$ should
remain relatively the same under the same local non-equilibrium
conditions when we change the size of the system. 
For systems in thermal gradients, one might expect 
\begin{equation}
  \label{d-loss}
  {\Delta  D\over D}{  = }C' \left(\nabla T\over T\right)^2
  +{\cal O}\left(\left(\nabla T\over T\right)^4\right)
\end{equation}
where the constant $C'$ might depend on $T$. Such a
behavior is particularly natural if one envisions that
a continuum limit of the theory ultimately exists.
$\Delta D$ has been studied previously for color
conduction\cite{hp1,hp3,cels}, sheared
fluids\cite{morriss,hp2,evans} and thermal conduction\cite{hp2},
numerically. Analytic computations of the Lyapunov spectrum and
$\Delta D$ have understandably been restricted to small or idealized
systems such as the Lorentz gas\cite{l-gas,cels,analy}.  The physical
properties are far from trivial; even whether $\Delta D$ generally
arises has been an issue\cite{fractal-controversy}.
Extensivity of
$\Delta D$ under thermal gradients has been analyzed
\cite{hp2}, but the relation to transport and entropy
production was not elucidated previously. Extensivity has been
investigated in  sheared fluids\cite{hp3,evans} and for color
conductivity\cite{hp1,hp3}.  Study of the dependence
on the number and types of thermostats or systematic analysis far
from equilibrium have not been performed before.

We first discuss the properties of $\Delta D$, its extensivity
and its relation to transport. Consider a system with cross-sectional
area $A$ placed in contact with
two heat-baths at both ends having temperatures
$(T_1^0,T_2^0)$. The Lyapunov spectrum $\left\{\lam
  j\right\}_{j=1,2,\ldots,D}$ distills the microscopic
properties of the classical system and are the time averaged
eigenvalues of the stability matrix $(\partial \dot
\varphi_i/\partial \varphi_j)$, where
$\left\{\varphi_j\right\}_{j=1,2,\ldots,D}$ are all the degrees
of the freedom of the system, {\it including} the heat-baths.
{}From the spectrum, we may compute the fractal dimension using the
Kaplan--Yorke estimate $\dky$, which has the
property that $\Delta D\equiv D-\dky>0$ in non-equilibrium and
$\Delta D=0$ in equilibrium systems.
While the extensivity relation \eqnn{d-loss} is natural, the
relation is ill-defined for two reasons: First, $\nabla T$
is in general {\it not} constant within the system and which
$\nabla T$ we choose visibly affects the results\cite{ak-long}.
Second, it is
unclear what $D$ in the 
denominator should mean. It can be the number of degrees of
freedom of the whole system including the thermostats, that
without, or that only of the interior.
These difficulties arise since $\dky$ is a global
quantity defined only for the whole system.
An expression consistent with
\eqnn{d-loss} in the near equilibrium limit that suffers no such
ambiguity even far from equilibrium is (denoting the heat flux
by $J$),
\begin{equation}
  \label{dj}
  {\Delta  D\over V_{in}} = C_D J^2
\end{equation}
since  $J$ is constant throughout the system. $V_{in}$
is the interior volume of the  system.
We shall derive this relation near equilibrium and further verify
it numerically.

Computing $\Delta D$ requires the full Lyapunov spectrum, which
involves evolving $D^2+D$ degrees of freedom. This apparently
makes it difficult to explicitly check extensivity. Close to
equilibrium, this difficulty can be overcome as
follows\cite{hp1,evans}. Define $\dmax,\dmin$ as
\begin{equation}
  \label{dky1}
  \dmax\equiv-{\sum_{j=1}^D\lam j\over \lam{max}},\qquad
  \dmin\equiv{\sum_{j=1}^D\lam j\over \lam{min}}
\end{equation}
where $\lam{max},\lam{min}$ are the maximum and minimum
exponents. When $\Delta D\leq1$, $\dmin=\Delta D$ holds {\it
  exactly}.  (More generally, when $K-1<\Delta D\leq 
K$, we need only to compute $\sum^D_{j=1}\lam j$ and the $K$
lowest Lyapunov exponents to obtain $\Delta D$.) 
This holds when the system is close to equilibrium. 
$\nabla T/T\sim1/L$ and $\Delta D\sim L^{d-2}$
 (in $d$-dimensions)
so that  $\Delta D$   is always  small for large systems in 1-d.
%This is always  the case for large systems, since $\nabla
%T/T\sim1/L$, so that $\Delta D\sim1/L$.   
Since
$\lam{min}=-\lam{max}$ in equilibrium, $\dmax$ should also be a
good approximation to $\Delta D$, close to equilibrium. We 
analyze these conditions systematically in the following and make
these conditions more precise. $\sum_{j=1}^D\lam j$, 
the total rate of phase space contraction,
can be computed from the equations of motion of $\{\varphi_j\}$
alone and $\lam{max}$ can be computed
from evolving one tangent vector, so we need only to evolve
$2D$ degrees of freedom to compute $\dmax$ --- a huge reduction
for large $D$.

We now systematically investigate how the extensivity of $\Delta
D$ arises. First note that $\sum_{j=1}^D\lam j$ is the rate of
entropy production, $\dot S$, for the
system\cite{attractor,hp1}. We  use this to derive a
thermodynamic relation\cite{ak-long}
\begin{eqnarray}
  \label{sumJ}
  \sum_{j=1}^D\lam j &=& -\dot S=AJ\left({1\over T_1^0}-{1\over
      T_2^0}\right)\nonumber\\
  &=&{V_{in}J^2 \over \kappa T^2}
  \left(1+{2\alpha\kappa\over V_{in}}\right) + {\cal O}(J^4)
\end{eqnarray}
where $\kappa$ is the thermal conductivity. We have included the
effects of boundary temperature jumps that behave as $\alpha J$
when the jumps are not too big\cite{ak-jumps}. $\alpha$, which
arises as a finite size correction, measures the efficacy of the
heat-baths, which can be stochastic or deterministic,
and can have significant effects as will be shown. Notice that (4)
always holds, both close to and far from equilibrium and is
independent of the type and number of thermostats used. Further,
it is $V_{in}$ that arises in the relation, rather than $D$.

$\lam{max}$ behaves as 
$  \lam{max}=\lam{max}^{eq}+{\cal O}(J^2)$,
where $\lam{max}^{eq}$ is its equilibrium value.
$\lam{max}^{eq}$ is {\it independent} of $V_{in}$ for large
enough systems, but can depend on $T$ and also on the
thermostats used.  Close to equilibrium, the above behavior of
$\sum_{j=1}^D\lam j$ and $\lam{max}^{eq}$, when combined,
explain the extensivity of $\Delta D$ as in \eqnn{dj}.  In this
limit, the  extensivity of $\Delta D$ arises from
$\sum_{j=1}^D\lam j$ and the
thermostat dependence from $\lam{max}^{eq}$.  Since $C_D$ can be
derived in the $J\rightarrow0$ limit, we derive
\begin{equation}
  \label{dj1}
  C_D=    {1\over\kappa\lam{max}^{eq}T^2}
  \left(1+{2\alpha\kappa\over V_{in}}\right)
  {{V_{in}\rightarrow\infty}\atop \longrightarrow}
  {1\over\kappa\lam{max}^{eq}T^2}
\end{equation}
which relates macroscopic transport and entropy production to the
microscopic $\Delta D$.  A subtlety needs to be mentioned: We have
found that $\lam{max}^{eq}$ is consistent with having a finite
large volume limit but have not proven this statement
analytically. In fact, this difficult issue is open and there have
been conflicting results on the existence of the thermodynamic
limit of $\lam{max}^{eq}$\cite{lmax}. From our argument, we see that
the extensive nature of $\Delta D$  requires that $\lam{max}^{eq}$ have
a thermodynamic limit or vice versa. We note that in systems
without a bulk limit, such as the FPU model, $\Delta D$ needs not
be extensive and $\lam{max}^{eq}$ needs not have a thermodynamic limit,
which might explain some of the discrepancy seen in the previous
literature\cite{lmax}.  Far from equilibrium, $\Delta D$ and
$\dmax$ will be quite different, yet the extensivity relation
\eqnn{dj} can and seems to still hold for $\Delta D$. This
reflects the deeper geometric significance of $\Delta D$ not
present in $\dmax$.

While the above theory is valid in any dimension, we now apply
it to the 1--d $\phi^4$ theory described by the following
Hamiltonian:
\begin{equation}
  \label{ham}
  H =\sum_{x=1}^L\left[{\pi_{x}^2\over2} +
    {\left(\nabla \phi_{x}\right)^2\over2}
    + {\phi_x^4\over4}
  \right]\quad.
\end{equation}
We choose the $\phi^4$ theory because it is a classic
statistical model that naturally appears in broad physical
contexts.
Also, the statistical properties of the theory have been studied
previously, including thermal transport which has bulk
behavior\cite{ak-long,hu-b}.  We model the heat-baths dynamically
by applying Nos\'e-Hoover (NH) thermostats or their variants
(demons) at the boundaries, as explained in \cite{ak-long}. The
interior includes the dynamics only of the $\phi^4$ theory.
Temperature is defined unambiguously using the ideal gas
thermometer, $T(x)=\langle \pi_x^2\rangle$.  The number of
(heat-bath) boundary sites thermostatted on each end, $N_B$, will
be varied. We employ one set of thermostats per thermostatted
site.  We run the simulations with time steps of $10^{-3}$ to
$0.05$ for $10^6$ to $10^9$ steps, understanding its convergence
properties and also checking that the results do not change with
the step size. To obtain the Lyapunov spectrum, we use the method
%explained in \cite{hp1}.
in \cite{lyapunovAlgorithm,hp1}.

\begin{figure}[htbp]
  \begin{center}
    \leavevmode
    \epsfxsize=86mm \epsfbox{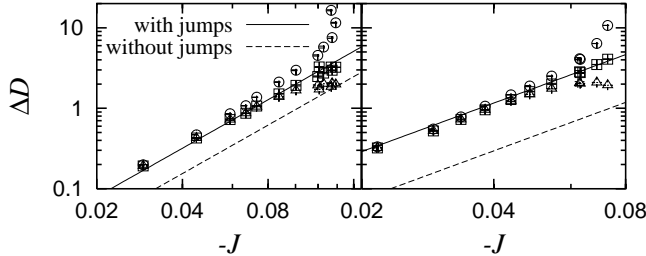}
    \caption{
      $\Delta D\ (\Box),\ \dmax\ (\circ)$ and $\dmin\
      (\bigtriangleup)$ against $-J$ in $L=11$ (left),
      $L=41$ (right) at  $T=0.5$ for the $N_B=1$ case.
      $J^2$ behavior with and without finite size corrections are also
      shown.  $\Delta D$ displays $J^2$ behavior even for large
      $-J$.
      }
    \label{fig:dloss1}
  \end{center}
\end{figure}

Let us first look at the dependence of $\Delta D$, $\dmax$ and
$\dmin$ with respect to $J$ (\figno{dloss1}). We see that close
enough to equilibrium, all three quantities agree and display
$J^2$ behavior as in \eqnn{dj} and \eqnn{dj1}. Remarkably, even
far from
equilibrium and well into the non--linear regime, $\Delta D$ has
a robust $J^2$ behavior, but not $\Delta D_{max,min}$. We
further note that the boundary temperature jumps have a
significant effect, the effect being larger for smaller $T$.
For instance, using
$\kappa=2.83(4)T^{-1.35(2)},\alpha=2.6(1)$\cite{ak-long},
${2\alpha\kappa}=300,6$ for $T=0.1,2$, respectively.

\begin{figure}[htbp]
  \begin{center}
    \leavevmode
    \epsfxsize=86mm \epsfbox{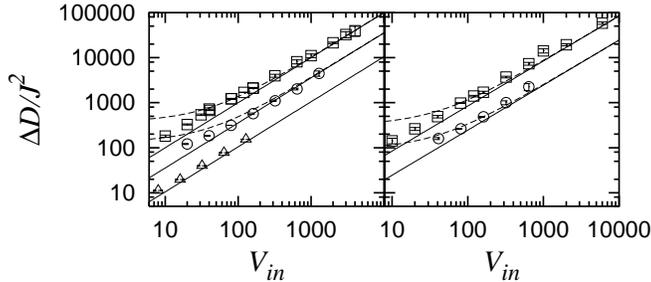}
    \caption{
      (left) $\Delta D/J^2$ against $V_{in}$ for 
     $N_B=1\ (\Box)$, $N_B=40\ (\circ)$,  with NH thermostats
     (left) and demons (right) at $T=0.5$.
     $V_{in}/(\kappa\lam{max}^{eq}T^2)$ (solid)        and
      $V_{in}/(\kappa\lam{max}^{eq}T^2)(1+2\alpha\kappa/V_{in})$
      (dashes) are plotted. Application of the formulas to the
     data of [4] ($\bigtriangleup$) works very well (data 
     and formulas rescaled by 5000 for plotting).
     }
    \label{fig:dloss2}
  \end{center}
\end{figure}

Performing an analysis similar to those performed for
\figno{dloss1}, we can extract the proportionality constant
$\Delta D/J^2$ for a particular $L$, $N_B$ and $T$. Combining
this data for various $V_{in}(=L-2N_B+1)$, $N_B$, we find that
the relations \eqnn{dj},\eqnn{dj1} describe the results quite
well over a few orders in magnitude (\figno{dloss2}). 
We have also included the data of \cite{hp2} which used {\it
  stochastic} thermostats and see that the formulas work
quite well.

There are a few subtleties which we now resolve: First, we can
and have made the distinction between the total volume and
$V_{in}$ in the formula \eqnn{dj}, since we have performed
analyses with $N_B=1\sim40$ sites. Secondly, $C_D$ can and does
depend on $T$ and also on the type of thermostats used,
including $N_B$.
$C_D$ has similar behavior both for NH
thermostats and for demons, with the former being larger, and
both decreasing with  $N_B$.
Since the demons have twice the number of degrees of
freedom per thermostatted site compared to NH thermostats, this
is quite remarkable. Furthermore, somewhat surprisingly, when we
increase $N_B$ by one, thereby {\it   increasing} the total
number of degrees of freedom by 6(8) 
in the NH thermostat (demons) case, $C_D$
{\it decreases} and so does $\Delta D$ for the same $J$. The
reason for this will be clarified below.

\begin{figure}[htbp]
  \begin{center}
    \leavevmode
    \epsfysize=32mm \epsfbox{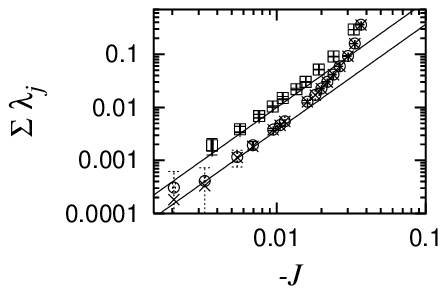}\hspace{2mm}
    \epsfysize=32mm \epsfbox{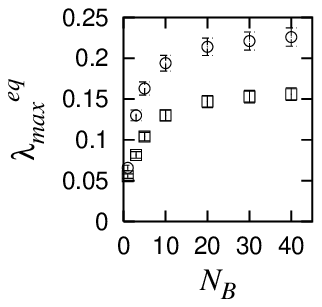}
    \caption{(left) $\sum_{j=1}^D\lam j$ and
      $J\left({1/ T_1^0}-{1/ T_2^0}\right)$ with respect to
      $-J$, for $T=0.5$ ($\Box,+$), $T=2$ ($\circ,\times$) and
      their quadratic behavior Eq.~\eqnn{sumJ} near equilibrium
      (solid) for the $L=162, N_B=1$ case.  The quantities agree
      excellently. \hfil\break
      (right) The dependence of the
      $\lam{max}^{eq}$ on $N_B$ at $T=0.5$ for NH
      thermostats ($\Box$) and demons ($\circ$). This demonstrates that
      $\lambda_{max}^{eq}$ is strongly heat-bath dependent, and is not
      uniquely determined by the local equilibrium conditions.}
    \label{fig:lmax}
  \end{center}
\end{figure}

We study examples of the behavior of $\sum_{j=1}^D\lam j$ in
\figno{lmax}~(left). We see that the entropy production relation
Eq.~\eqnn{sumJ} works well near and far from equilibrium,
and that the quadratic behavior with respect to $J$ can be observed
close to equilibrium. This is true, {\it   independently } of
the type of thermostats used. On the other hand, $\lam{max}^{eq}$ can
and does depend on the type of thermostats used as well as $T$
(\figno{lmax}~(right)). This is responsible for the thermostat
dependence of $C_D$ seen in \figno{dloss2}.
The larger
$\lam{max}^{eq}$ for demons and the increase seen with $N_B$ are
quite natural since the existence of more thermostats lead to
more chaotic behavior.

We now study in detail the difference between $\Delta D,$ $ \dmin$
and $\dmax$: Let us first work out what the condition $\Delta
D\leq1$ means.  Define $r\equiv\Delta T/T$, where $T$ is the
central temperature and $\Delta T$ is
the difference in the boundary temperatures so that $r\leq2$,
always, and $\nabla T\simeq rT/V_{in}$.  Then, we can derive a
simple estimate,
\begin{equation}
  \label{dleq1}
  \Delta D\leq1
  \quad\Leftrightarrow\quad
  V_{in}\geq {\kappa\left(\nabla T\right)^2\over
    \lam{max}^{eq}T^2} \simeq r^2 { \kappa\over\lam{max}^{eq}}
\end{equation}
Let us analyze $T=0.5$ case more concretely; the condition is
most stringent for $N_B=1$ (NH) case, when $\lam{max}^{eq}$ is
the smallest. So we obtain the condition $V_{in}/r^2\gtrsim130$
for $\Delta D\leq1$ and for large lattices, $V_{in}\gtrsim400$,
it should be satisfied for any gradient, which is consistent with
our results.
While we
did not consider here the non-linearity of the profiles, the
boundaries temperature jumps and the non-linearity of the
response\cite{ak-jumps,ak-le},
we are in some sense close to equilibrium when $\Delta
D\leq1$ so that the rough arguments suffice for the purpose at hand.

When $\Delta D\leq1$, how large is the difference $\dmax-\Delta
D$?  In this case, $\dmin=\Delta D$, as explained above, so 
we need to consider $\lam{max}+\lam{min}$, which is zero
in equilibrium since the
Lyapunov spectrum is symmetric with respect to sign
inversion\cite{attractor}.  As we move away from equilibrium, the
behavior can be described by
${\lam{max}/\lam{min}}+1=
C_\lambda J^2+{\cal O}\left(J^4\right) $.
Then,  when $\Delta D\leq1$,
\begin{equation}
  \label{error-dmax}
  {\dmax-\Delta D\over\Delta D}\simeq C_\lambda J^2
  \leq {C_\lambda\over V_{in}C_D}
\end{equation}
For $N_B=1$, $T=0.5$, $C_\lambda=13 L^{0.5}$, so that
${(\dmax-\Delta D)/\Delta D} \lesssim
0.1\times(V_{in}/100)^{-0.5}$.  Since the statistical errors
are typically at a few \% level, the difference between $\Delta
D$ and $\dmax$ is at most comparable to them except for small
systems, when $\Delta
D\leq1$. In  \figno{dloss1}, it can indeed be seen that for $\Delta
D\leq1$, $\dmax$ agrees with $\Delta D$ within error.
Similar analysis can be applied
at different $T$.

When $\Delta D\geq1$, the relation $\dmin=\Delta D$ no longer
holds exactly and the deviations from it is of ${\cal
  O}(\delta\lambda/\lambda)$, with $\delta\lambda$ being the
sum of the differences of the exponents $\lam j\ (j> \dky)$, which
is {\it not} zero even in equilibrium.
$\delta\lambda/\lambda\sim\Delta D/D$ so  that in systems with
small relative dimensional loss, $\dmax$ can still be a good
approximation.
In \figno{dloss1}, we see that when $\Delta D\gtrsim1$, $\dmax$
is in general not a good approximation to $\Delta D$ and that
it is a clearly better approximation  there for the larger system.

We derived the extensivity of $\Delta D$ in systems with bulk
behavior under thermal gradients in \eqnn{dj}, thereby making the
notion precise. The extensivity was explicitly related to the
macroscopic transport properties as in \eqnn{dj1} through entropy
production.  Previously, it was emphasized that the extensivity of
$\Delta D$ is not compatible with local equilibrium so that it is
questionable\cite{cels}. It is now known that systems such as
$\phi^4$ theory in $d=1-3$ display violations of local equilibrium
under thermal gradients in a similar manner, as $\sim\left(\nabla
T/T\right)^2$ \cite{ak-le}. This resolves the apparent conflict
since the violations of local equilibrium emerge in a manner
analogous to (2).

We further explicitly verified that $\Delta D$ in $\phi^4$
theory behaves extensively under various thermal gradients for
$L\lesssim10^4$ and its relation to transport. The
relations \eqnn{dj},\eqnn{sumJ} and \eqnn{dj1}, however, are
more general.  
We saw that the relations applied well to 
\cite{hp2} which used {\it stochastic} thermostats in a
different model. 
An application to dilute gas using the standard estimates of
$\lam{max}$\cite{attractor} yields $C_D\simeq2/[\rho
v^2\ln(4l/d)]$, where $\rho$ is the density, $v$ is the
average particle velocity, $l$ is the mean free path and $d$ is
the particle diameter. Then for $\nabla T/T\sim0.01\,$m$^{-1}$, 
$\Delta D\sim10^8$\,${\rm m}^{-3}$ at room temperature ---
quite large, yet 
far smaller than the total number of degrees of freedom.

We find the results satisfying from the physics point of view:
Since $\Delta D$ pertains to the whole system, it includes the
temperature profile which is curved in general, boundary
temperature jumps and
the various types of thermostats. Yet, $\Delta D $ can be related
to macroscopic transport with the thermostat dependence cleanly
separated into $\lam{max}$. Furthermore, $\Delta D$ has extensive
behavior with respect to the internal volume wherein the
system is manifestly in non-equilibrium. 
We have seen that
$\lam{max}^{eq}$ is {\sl not} unique: In global thermal
equilibrium, different choices of heat-baths can lead to very
different values. The result is that dimensional loss is not
unique either, only the product  $\Delta D\lam{max}^{eq}$ 
behaves macroscopically and can be 
related to thermodynamic quantities.

\end{document}